\documentclass[floatfix,showkeys]{revtex4}
\usepackage{amsmath,amssymb}
\usepackage[dvips]{graphicx}
\usepackage[dvips]{color}
\usepackage{mathrsfs}

\newcommand{\be}{\begin{equation}}
\newcommand{\ee}{\end{equation}}
\newcommand{\nablab}{\boldsymbol{\nabla}}
\newcommand{\IF}{\mathscr{I}}

\newcommand{\Tu}{T_{\!\boldsymbol{u}}}
\newcommand{\uu}{\boldsymbol{u}}

\newcommand{\rr}{\boldsymbol{r}}

\newcommand{\Pu}{\boldsymbol{P}_{\!\!\boldsymbol{u}}}

\newcommand{\Pc}{\boldsymbol{P}_{\!\!c}}

\newcommand{\um}{\textrm{one}}
\newcommand{\ro}{\rho}
\newcommand{\p}{p}
\newcommand{\pn}{p_{\scriptscriptstyle N}}

\newcommand{\eq}[1]{Eq.~(\ref{#1})}
\newcommand{\eqs}[1]{Eqs.~(\ref{#1})}

\begin{document}

\title{Classical kinetic energy, quantum fluctuation terms and kinetic-energy functionals}

\author{I. P. Hamilton}
\email{ihamilton@wlu.ca}\affiliation{Department of Chemistry,
Wilfrid Laurier University, Waterloo, Canada N2L 3C5.}

\author{Ricardo A. Mosna}
\email{mosna@ime.unicamp.br} \affiliation{ Instituto de
Matem\'atica, Estat\'\i stica e Computa\c{c}\~ao Cient\'\i fica,
Universidade Estadual de Campinas, C.P. 6065, 13083-859, Campinas,
SP, Brazil.}

\author{L. Delle Site}
\email{dellsite@mpip-mainz.mpg.de} \affiliation{Max-Planck-Institute
for Polymer Research, Ackermannweg 10, D 55021 Mainz Germany.}


\begin{abstract}

We employ a recently formulated dequantization procedure to obtain
an exact expression for the kinetic energy which is applicable to
all kinetic-energy functionals. We express the kinetic energy of an
$N$-electron system as the sum of an $N$-electron classical kinetic
energy and an $N$-electron purely quantum kinetic energy arising
from the quantum fluctuations that turn the classical momentum into
the quantum momentum. This leads to an interesting analogy with
Nelson's stochastic approach to quantum mechanics, which we use to
conceptually clarify the physical nature of part of the
kinetic-energy functional in terms of statistical fluctuations and
in direct correspondence with Fisher Information Theory. We show
that the $N$-electron purely quantum kinetic energy can be written
as the sum of the (one-electron) Weizs\"{a}cker term and an
($N$-1)-electron kinetic correlation term. We further show that the
Weizs\"{a}cker term results from local fluctuations while the
kinetic correlation term results from the nonlocal fluctuations. We
then write the $N$-electron classical kinetic energy as the sum of
the (one-electron) classical kinetic energy and another
($N$-1)-electron kinetic correlation term. For one-electron orbitals
(where kinetic correlation is neglected) we obtain an exact (albeit
impractical) expression for the noninteracting kinetic energy as the
sum of the classical kinetic energy and the Weizs\"{a}cker term. The
classical kinetic energy is seen to be explicitly dependent on the
electron phase and this has implications for the development of
accurate orbital-free kinetic-energy functionals. Also, there is a
direct connection between the classical kinetic energy and the
angular momentum and, across a row of the periodic table, the
classical kinetic energy component of the noninteracting kinetic
energy generally increases as $Z$ increases. Finally, we underline
that, although our aim in this paper is conceptual rather than
practical, our results are potentially useful for the construction
of improved kinetic-energy functionals.

\end{abstract}

\maketitle

\section{Introduction}
\label{sec:intro}

\subsection{Density Functional Theory}
Density functional theory has developed into an extremely successful
approach for the calculation of atomic and molecular properties.
\cite{YangParr,DG,KH} In this approach, the electron density,
$\ro(\rr)$, is the fundamental variable and properties such as the
energy are obtained from $\ro$ rather than from the $N$-electron
wavefunction, $\psi(\rr_1,\ldots,\rr_N)$, as in conventional quantum
mechanical approaches based on the Schr\"odinger equation. The
motivation for density functional theory is clear --- if properties
such as the energy can be obtained from $\ro$ then calculations on
systems with a large number of electrons are, in principle, no more
difficult than those on systems with a small number. However, this
depends on having accurate energy functionals which, in practice, is
a serious problem. The energy can be partitioned into kinetic and
potential terms and a clear zeroth-order choice of functional for
the potential energy is the classical expression $-Ze^2\int \frac
{\ro(\rr)}{r}d^3\rr + \frac{e^2}{2}\int \int \frac
{\ro(\rr_1)\ro(\rr_2)}{r_{12}}d^3\rr_1d^3\rr_2$. However, for atomic
and molecular systems, there is no correspondingly clear
zeroth-order choice of functional for the kinetic energy.

\subsection{Quantum fluctuations}

One of the key aspects of quantum mechanics is that one cannot
simultaneously ascribe well-defined (sharp) values for the position
and momentum of a physical system. Motivated by this, quantization
procedures have been proposed in which the quantum regime is
obtained from the classical regime by adding stochastic terms to the
classical equations of motion. In particular, Nelson \cite{nelson}
and earlier work of F\'enyes \cite{FE} and Weizel \cite{WE} has
shown that the Schr\"odinger equation can be derived from Newtonian
mechanics via the assumption that particles are subjected to
Brownian motion with a real diffusion coefficient. The Brownian
motion results in an osmotic momentum and adding this term to the
classical momentum results in the quantum momentum.

We recently proposed \cite{MHD06} a dequantization procedure whereby
the classical regime is obtained from the quantum regime by
stripping these ``quantum fluctuations'' from the quantum momentum
resulting in the classical momentum. In particular, we introduced
deformations of the momentum operator, which correspond to generic
fluctuations of the particle's momentum. These lead to a deformed
kinetic energy, which roughly quantifies the amount of ``fuzziness''
caused by these fluctuations. We showed that the deformed kinetic
energy possesses a unique minimum, which is seen to be the classical
kinetic energy. In this way, a variational procedure determines the
particular deformation that has the effect of suppressing the
quantum fluctuations, resulting in dequantization of the system.
From this variational procedure we obtain a term (identical to the
osmotic momentum of Nelson \cite{nelson}) which, when added to the
classical momentum results in the quantum momentum. This is an
interesting point which is further clarified in this paper: the
classical limit of the physics of electrons, with its usual
statistical interpretation, finds a direct correspondence with our
dequantization procedure as we show later.

In this paper we obtain an expression of the quantum-classical
correspondence for the kinetic energy when $\ro$ is the fundamental
variable for the quantum terms. In this expression the kinetic
energy of an $N$-electron system is written as the sum of an
$N$-electron classical kinetic energy and an $N$-electron purely
quantum kinetic energy arising from the quantum fluctuations that
turn the classical momentum into the quantum momentum, as in Nelson's
stochastic approach to quantum mechanics. \cite{nelson} We establish
a connection between the osmotic momentum of Nelson, the
Weizs\"{a}cker term and the Fisher Information. For one-electron
orbitals we then obtain an expression for the noninteracting kinetic
energy as the sum of the classical kinetic energy and the
Weizs\"{a}cker term. The Weizs\"{a}cker term is well-known and the
classical kinetic energy is related to the Thomas-Fermi term which
is also well-known. However, we believe that our derivation, which
obtains {\it both} these terms within a {\it single} theoretical
framework, is novel. Also, there are significant differences between
the classical kinetic energy and the Thomas-Fermi term. In
particular, the classical kinetic energy is explicitly dependent on
the electron phase. Our expression is therefore at best order $N^3$
and can have no practical advantage over the standard Kohn-Sham
expression. However, our expression is exact and we will show that
it correctly reduces to the Thomas-Fermi term for the uniform
electron gas and to the Weizs\"{a}cker term for the hydrogen atom.
By examining our expression for basis functions that are the product
of radial functions and spherical harmonics, we establish a direct
connection between the classical kinetic energy and the angular
momentum. We believe that this intrinsic connection between the
angular momentum and a component of the noninteracting kinetic
energy is of significant conceptual value in showing the information
that should be incorporated in any kinetic-energy functional.

\section{Kinetic-energy functionals}

We begin by considering some previously proposed kinetic-energy
functionals whereby the kinetic energy is obtained from the electron
density, $\rho$. Here the electron density is given in terms of the
(normalized) wavefunction by \be
\rho(\rr)=N\int|\psi(\rr,\ldots,\rr_N)|^2 \, d^3\rr_2\ldots
d^3\rr_N, \ee so that $\int \rho(\rr)\,d^3\rr = N$.

\subsection{Thomas-Fermi and Weizs\"{a}cker terms}
A well-known functional for the kinetic energy, formulated by
Thomas and Fermi \cite{thomas,fermi}, is
\begin{equation}
T_{TF}=\frac{3\hbar^2}{10m}(3\pi^2)^{2/3} \int \ro(\rr)^{5/3} d^3\rr.
\label{TF}
\end{equation}
This expression is exact for the
uniform electron gas (an $N =\infty$ system) for which the reduced
gradient ($|\nablab\ro|/2k_f\ro$ with $k_f = (3\pi^2\ro)^{1/3}$) is
zero. Another well-known kinetic-energy functional, formulated by
Weizs\"{a}cker \cite{weizsacker}, is

\begin{equation}
T_{W}=\frac{\hbar^2}{8m}\int\frac{|\nablab\ro(\rr)|^2}{\ro(\rr)}d^3\rr.
\label{TW}
\end{equation}
This expression is exact for the ground state of the hydrogen atom
(an $N = 1$ system).

For atomic systems it might be hoped that an accurate kinetic energy
functional could be obtained via some combination of $T_{TF}$ and
$T_W$ and, in fact, Weizs\"{a}cker had proposed $T_{TF} + T_W$.
Other researchers subsequently proposed either a smaller coefficient
for $T_{TF}$ \cite{MY,absp,GL,gr,acharya} or, more commonly, $T_W$.
A second-order gradient expansion of the density for a nonuniform
electron gas (valid for small reduced gradient) leads to the
coefficient $\frac{1}{9}$. \cite{KP,kirzhnits,yang} Other
expressions for the kinetic energy have been developed and, of
particular relevance to our paper, Herring \cite{herring} proposes
$T_\theta$ + $T_W$ where $T_\theta$ is termed the relative-phase
energy. In our expression for the kinetic energy the relative-phase
energy is replaced by the classical kinetic energy.

For large $Z$ atoms, the electron density is slowly varying for the
bulk of the electrons in the intermediate $r$ region, a second-order
gradient expansion is valid, and the expression $T_{TF}$ +
$\frac{1}{9}$$T_W$ (with the Dirac exchange functional \cite{dirac})
is accurate. However, this expression is not accurate for small and
large $r$. \cite{Spruch} For small $r$ the Scott correction,
\cite{Scott} can be employed but for large $r$ no correction is
known. Unfortunately, the large $r$ region is (by virtue of the
valence electrons) responsible for chemical bonding and Thomas-Fermi
theory cannot describe molecular systems. An expression for the
kinetic energy which is accurate for large $r$ and which might, in
principle, be employed to correct the Thomas-Fermi expression in the
large $r$ region would therefore be of significant interest.

The Fisher information, \cite{fisher,nagy} which was developed
in information theory as a measure of spatial localization, is given by
\begin{equation}
\IF= \int \frac{|\nablab\p(\rr)|^2}{\p(\rr)} d^3\rr,
\label{FI}
\end{equation}
where $\p(\rr_1)=\int|\psi(\rr_1,\ldots,\rr_N)|^2 \, d^3\rr_2\ldots
d^3\rr_N$ is the one-electron (probability) density, so that
$\rho(\rr)=N\p(\rr)$. It follows that $T_{W}=\frac{N\hbar^2}{8m}\IF$
and these quantities are essentially identical.

\subsection{Hohenberg-Kohn theorems and Kohn-Sham approach}
\label{sec:hkks}

Density functional theory was placed on a solid foundation by the
work of Hohenberg and Kohn \cite{HK} who proved that the total
energy can indeed be obtained as a functional of $\ro$. Their proof
also applies to the kinetic energy but they could provide no
prescription for constructing the exact kinetic-energy functional.
Kohn and Sham \cite{KS} subsequently provided a prescription for
calculating the noninteracting kinetic energy by adapting aspects of
Hartree-Fock theory. In Hartree-Fock theory the wavefunction is
approximated as the product of $N$ one-electron orbitals
(antisymmetrized to ensure that electron exchange is incorporated
exactly for the approximate wavefunction). In constructing these
orbitals the effect of the other electrons is included only in an
average way (through the use of an effective potential) and electron
correlation is neglected. Calculations scale as $N^3$ and post
Hartree-Fock approaches incorporating electron correlation (required
for chemical accuracy) typically scale as $N^5$ or $N^7$. Kohn and
Sham employed the orbital approximation but chose the effective
potential such that for the one-electron orbitals, $\phi_i$, the
resulting density is equal to $\ro$. From these orbitals they
obtained the noninteracting kinetic energy as $T_s = \frac
{\hbar^2}{2m}\int \sum_{i=1}^N |\nablab\phi_i|^2 d^3\rr$ rather than
as a direct functional of $\ro$. As in Hartree-Fock theory, electron
exchange is incorporated exactly and electron correlation is
neglected. Complete calculations employ an exchange-correlation
functional for the difference between $T_s$ and the exact kinetic
energy (and also the difference between the classical electrostatic
energy and the exact potential energy). In the canonical
implementation (with semi-local approximations to the
exchange-correlation potential) calculations scale as $N^3$ as in
Hartree-Fock theory but, because high-quality exchange-correlation
functionals have been developed, chemical accuracy can be realized
and it is in this form that density functional theory has been most
successful for the calculation of atomic and molecular properties.

Despite the success of the Kohn-Sham approach, there has been
continued interest in developing expressions (termed orbital-free
kinetic-energy functionals) which obtain the noninteracting kinetic
energy, $T_s$, as a direct functional of $\ro$. The very practical
motivation is that these expressions could be order $N$ and much
larger systems would therefore be tractable but chemical accuracy
has not yet been realized. A recent study \cite{CCH} carefully
analyzed kinetic-energy functionals of the $T_{TF}$ + $\lambda T_W$
form while other recent studies \cite{IEMS,TW} considered the
accuracy of various kinetic-energy functionals which combine
$T_{TF}$, $T_W$ and higher-order gradient expansion terms in more
complicated ways. The development of orbital-free kinetic-energy
functionals continues to be an active area of research.
\cite{SLBB,JY,CW,BC,ON,ZW}

\section{Quantum-classical correspondence}
\label{sec:dequant}

Consider, for an $N$-electron system, a local deformation
$\boldsymbol{P}\to\Pu$ of the quantum momentum operator
$\boldsymbol{P}=-i\hbar\nablab$, with \cite{MHD06}
\begin{equation}
\Pu\psi = \left( \boldsymbol{P} - i\uu \right) \psi,
\label{Pu}
\end{equation}
where all quantities in bold face are 3$N$-dimensional vectors and
$\uu$ is real.

Let
\begin{equation}
T=\frac{1}{2m}\int (\boldsymbol{P}\psi)^\ast (\boldsymbol{P}\psi) d^{3N}\rr \label{Tqm}
\end{equation}
and
\begin{equation}
\Tu =\frac{1}{2m}\int (\Pu\psi)^\ast (\Pu\psi) d^{3N}\rr \label{Th0}
\end{equation}
be the kinetic terms arising from $\boldsymbol{P}$ and $\Pu$,
respectively.

We recently showed \cite{MHD06} that extremization of $\Tu$ with
respect to $\boldsymbol{u}$-variations leads to the critical point
\begin{equation}
\uu_c = -\frac{\hbar}{2}\frac{\nablab\pn}{\pn}, \label{umin}
\end{equation}
where $\pn(\rr_1,\ldots,\rr_N)=|\psi(\rr_1,\ldots,\rr_N)|^2$ is the
$N$-electron (probability) density (with $\int \pn d^3\rr_1\cdots
d^3\rr_N$ = 1). We previously \cite{MHD05} obtained the same
expression for $\uu_c$ via a Witten deformation of the quantum
momentum. This value of $\uu_c$ results in the classical momentum
operator \cite{MHD06,MHD05}
\begin{equation}
\Pc \psi = \left( \boldsymbol{P} +
\frac{i\hbar}{2}\frac{\nablab\pn}{\pn} \right) \psi.
\label{Pc}
\end{equation}
Thus our dequantization procedure \emph{automatically} identifies
the expression for $\uu_c$ (cf \eq{umin}) which when added to the
quantum momentum results in the classical momentum. Here $-\uu_c$ is
identical to the osmotic momentum of Nelson \cite{nelson}, and
adding $-\uu_c$ to the classical momentum results in the quantum
momentum.

This value of $\uu_c$ results in
\be
T_{\uu_c}=T-\frac{\hbar^2}{8m}\IF_N,
\label{Tumin}
\ee
where $\IF_N$ is the $N$-electron Fisher information \cite{fisher}
\be
\IF_N= \int \frac{\left( \nablab\pn \right)^2}{\pn} d^{3N}\rr.
\label{fisher}
\ee

If the wavefunction is written as $\psi=\sqrt{\pn}e^{iS_N/\hbar}$
where $S_N(\rr_1,\ldots,\rr_N)$ is the $N$-electron phase then a
straightforward calculation shows that the action of $\Pc$ on $\psi$
is given by
\begin{equation}
\Pc\psi=\nablab S_N \: \psi, \label{Pcl1}
\end{equation}
so that, from \eq{Th0},
\begin{equation}
T_{\uu_c}=\frac{1}{2m} \int \pn \, |\nablab S_N|^2 d^{3N}\rr.
\label{Tuc2}
\end{equation}
This quantity is the mean kinetic energy of a classical ensemble,
described by the density $\pn$ and momentum $\nablab S_N$
\cite{Goldstein,Holland} and we therefore refer to $T_{\uu_c}$ as
the $N$-electron classical kinetic energy $T_{Cl,N}$.

\section{Results and Discussion}
\label{sec:decomp}

The $N$-electron kinetic energy can be expressed, from \eq{Tumin},
as
\begin{equation}
T_N = T_{Cl,N} + \frac{\hbar^2}{8m}\IF_N. \label{dec_T0}
\end{equation}
This is the sum of the $N$-electron classical kinetic energy and a
purely quantum term which is essentially given by the $N$-electron
Fisher information although, as our approach is restricted to scalar
particles, effects due to electron spin are not explicitly included
and our expressions are valid only for a single-spin wavefunction.

We first consider the $N$-electron classical kinetic energy of
\eq{dec_T0}. It immediately follows from \eq{Tuc2} that $T_{Cl,N}$=0
if and only if the $N$-electron phase is constant. Since a constant
$N$-electron phase can always be redefined to be zero, this is the
case if and only if the wavefunction is real.

We now consider the purely quantum term of \eq{dec_T0}. As in Ref.~\cite{sears}
we decompose the $N$-electron density as
\begin{equation}
\pn(\rr_1,\ldots,\rr_N)=\p(\rr_1)f(\rr_2,\ldots,\rr_N|\rr_1),
\label{eqfac}
\end{equation}
where
\[
\p(\rr_1)=\int \pn(\rr_1,\ldots,\rr_N) d^3\rr_2\cdots
d^3\rr_N
\]
and
\[
f(\rr_2,\ldots,\rr_N|\rr_1)=\frac{\pn(\rr_1,\ldots,\rr_N)}{\p(\rr_1)}.
\]
In this way, while $\p$ is the (already introduced) one-electron
probability density, the quantity $f(\rr_2,\ldots,\rr_N|\rr_1)$ is a
conditional density in that it represents the electron density
associated with a set of values for $\rr_2,\ldots,\rr_N$ given a
fixed value for $\rr_1$. Here $\p$ and $f$ satisfy the normalization
conditions
\begin{subequations}
\label{norma}
\begin{gather}
\int \p(\rr_1)d^3\rr_1 =1, \\
\int f(\rr_2,\ldots,\rr_N|\rr_1)d^3\rr_2\cdots d^3\rr_N =1 \;\;\forall\,\rr_1.
\end{gather}
\end{subequations}
This immediately yields an
expression for the minimizing momentum fluctuations (cf~\eq{umin}) as
\begin{equation}
-\uu_c=\frac{\hbar}{2}\left(
\frac{\nablab_{\rr_1}\rho(\rr_1)}{\rho(\rr_1)} +
\sum_{i=2}^N
\frac{\nablab_{\rr_i}f(\rr_2,\ldots,\rr_N|\rr_1)}{f(\rr_2,\ldots,\rr_N|\rr_1)}
\right),
\label{ueq}
\end{equation}
where the relation $\rho(\rr) = N\p(\rr)$ was used. In \eq{ueq} it
is implicitly assumed that $\rho$ and $f$ result from the same
wavefunction and that all necessary representability conditions are
therefore satisfied. From \eq{ueq} it is possible to distinguish a
{\em local} part of the momentum fluctuation,
$\frac{\hbar}{2}\frac{\nablab_{\rr_1}\rho(\rr_1)}{\rho(\rr_1)}$,
corresponding to fluctuation of the one-electron density in the
(arbitrary but fixed) variable $\rr_1$, and a {\em nonlocal} part,
$\frac{\hbar}{2}\sum_{i=2}^N
\frac{\nablab_{\rr_i}f(\rr_2,\ldots,\rr_N|\rr_1)}{f(\rr_2,\ldots,\rr_N|\rr_1)}$,
corresponding to fluctuation of the correlation function
$f(\rr_2,\ldots,\rr_N|\rr_1)$.

The $N$-electron Fisher information (cf \eq{fisher}) can be
written as
\[
\IF_N = N \int
\frac{\left[\nablab_{\rr_1}\pn(\rr_1,\ldots,\rr_N)\right]^2}{\pn(\rr_1,\ldots,\rr_N)}d^3\rr_1\cdots
d^3\rr_N.
\]
The decomposition for $\pn$ in \eq{eqfac} can then be used to
express this quantity in a more illuminating form as
\begin{align}
\IF_N &= N \int
\frac{\left[\nablab_{\rr_1}\p(\rr_1)f(\rr_2,\ldots,\rr_N|\rr_1)+\p(\rr_1)\nablab_{\rr_1}f(\rr_2,\ldots,\rr_N|\rr_1)\right]^2}{\p(\rr_1)f(\rr_2,\ldots,\rr_N|\rr_1)}
d^3\rr_1\cdots d^3\rr_N \notag\\
    &= N \int \frac{\left[\nablab_{\rr_1}\p(\rr_1)\right]^2}{\p(\rr_1)}d^3\rr_1+
       N \int\p(\rr_1)\frac{\left[\nablab_{\rr_1}f(\rr_2,\ldots,\rr_N|\rr_1)\right]^2}{f(\rr_2,\ldots,\rr_N|\rr_1)}d^3\rr_1\cdots d^3\rr_N,
\label{reduction}
\end{align}
where \eqs{norma} were used to simplify the first term and cancel the
mixed term. We then have
\be
\IF_N = \int
\frac{|\nablab\ro(\rr)|^2}{\ro(\rr)}d^3\rr+
  \int \ro(\rr) \IF_{\um}^{f}(\rr) d^3\rr,
\label{dec_fisher}
\ee
where
\[
\IF_{\um}^{f}(\rr) = \int
\frac{\left[\nablab_{\rr_1}f(\rr_2,\ldots,\rr_N|\rr)\right]^2}{f(\rr_2,\ldots,\rr_N|\rr)}d^3\rr_2\ldots d^3\rr_N.
\]
Thus Eq.~(\ref{dec_fisher}) decomposes the $N$-electron Fisher
information as a sum of two terms. The first is local, and is $N$
times $\IF$ (cf~\eq{FI}), and the second is nonlocal and comprises
many-electron effects through $\IF_{\um}^f$. This equation provides
a connection between the osmotic momentum of Nelson, the
Weizs\"{a}cker term and the Fisher Information.

\subsection{One-electron kinetic energy}

From Eqs.~(\ref{TW}), (\ref{dec_T0}) and (\ref{dec_fisher}), we
obtain the $N$-electron kinetic energy as
\begin{equation} T_N = T_{Cl,N} + T_W + \frac{\hbar^2}{8m}\int
\ro(\rr) \IF_{\um}^{f}(\rr) d^3\rr. \label{dec_T}
\end{equation}
Eq.~(\ref{dec_T}) decomposes the $N$-electron kinetic energy as the
sum of a classical term and two purely quantum terms and constitutes
an expression of the quantum-classical correspondence for the
$N$-electron kinetic energy when $\ro$ is the fundamental variable
for the quantum terms.

Each term of \eq{dec_T} adds an {\em independent} nonnegative
contribution to the kinetic energy and this equation agrees with the
decomposition of Sears et al. \cite{sears} when the $N$-electron
phase is constant (since $T_{Cl,N}$ is zero in this case, as
discussed above). Thus we see that the classical term in \eq{dec_T}
improves the lower bound for the general case in which the
$N$-electron phase is not constant.

In \eq{dec_T} the first quantum term contributes to the
noninteracting kinetic energy and the second contributes to the
kinetic correlation. We now assume that the $N$-electron classical
kinetic energy, $T_{Cl,N}$, can be decomposed as the sum of a term,
$T_{Cl}$, which contributes to the noninteracting classical kinetic
energy, and a term, $T_{Cl}^{corr}$, which contributes to the
classical kinetic correlation. Terms that contribute to the
noninteracting kinetic energy can be estimated by employing the
orbital approximation. If the one-electron orbital is written as
$\phi_i=\sqrt{\p}\,e^{iS_i/\hbar}$ where $S_i(\rr)$ is the electron
phase then $T_{Cl}=\frac{1}{2m} \int \p(\rr) \sum_{i=1}^N|\nablab
S_i(\rr)|^2 d^{3}\rr$ but we have no explicit expression for
$T_{Cl}^{corr}$. From Eq.~(\ref{dec_T}), we then obtain the
(one-electron) kinetic energy as
\begin{equation} T = T_{Cl} + T_{Cl}^{corr} + T_W + \frac{\hbar^2}{8m}\int
\ro(\rr) \IF_{\um}^{f}(\rr) d^3\rr. \label{dec_Tone}
\end{equation}

\subsection{Weizs\"{a}cker term, kinetic correlation term, and quantum fluctuations}

In Eq.~(\ref{dec_Tone}) the purely quantum terms, $T_W$ and
$\frac{\hbar^2}{8m}\int \ro(\rr) \IF_{\um}^{f}(\rr) d^3\rr$,
comprise the $N$-electron Weizs\"{a}cker term and, as discussed
above, arise in our approach from the fluctuations that turn the
classical momentum into the quantum momentum, as in Nelson's
stochastic approach to quantum mechanics. \cite{nelson} Many
decompositions of the $N$-electron Weizs\"{a}cker term are possible
\cite{miao,ayers} and, as noted above, a decomposition similar to
ours has previously been proposed \cite{sears}. The novelty of our
decomposition is that, from the calculation leading to
\eq{reduction}, we can unequivocally identify $T_W$ as resulting
from the local part of the quantum fluctuations, and
$\frac{\hbar^2}{8m}\int \ro(\rr) \IF_{\um}^{f}(\rr) d^3\rr$ as
resulting from the nonlocal part (cf \eq{ueq} and the discussion
following it). The latter term contributes to the kinetic
correlation and we note that an analytic expression for the electron
correlation which incorporates both kinetic and Coulombic terms has
been proposed. \cite{DS1} As noted above, $T_W$ (or $\IF$, which is
a measure of spatial localization) has been universally utilized to
construct kinetic-energy functionals and has also been employed to
characterize electronic properties \cite{gadre,romera1}. By also
employing the Shannon entropy power \cite{shannon}, which is a
measure of spatial delocalization, it has been possible to partially
characterize many-electron effects \cite{romera2,sagar}. However,
the connection between the kinetic correlation term and nonlocal
quantum fluctuations provides a new rationale for the need to
incorporate this term in exchange-correlation functionals in order
to capture the complete range of many-electron effects.

\subsection{Noninteracting kinetic energy}
\label{sec:noni}

In the orbital approximation kinetic correlation is neglected and
omitting these terms in Eq.~(\ref{dec_Tone}), we obtain the
noninteracting kinetic energy as
\begin{equation}
T_s = T_{Cl}+ T_W.
\label{dec_Ts}
\end{equation}
As we can see from the discussion preceding \eq{dec_Tone}, $T_{Cl}$
is linked to the electron phase and there are two limiting cases
where this expression is known analytically. For the ground state of
the hydrogen atom (an $N = 1$ system), the electron phase is zero,
so that $T_{Cl}=0$. Therefore, $T_s = T_W$ in this limit. For the
uniform electron gas (an $N=\infty$ system) the electrons are
noninteracting, so that \eq{dec_Ts} also applies. Since the
distribution is uniform, $T_W=0$ in this case and $T_{Cl} = T_s$ can
be calculated, as usual, by adding up the kinetic energies of
one-electron orbitals approximated as local plane waves, which
results in the Thomas-Fermi term. \cite{yang}

The standard expression for the noninteracting kinetic energy (see
section \ref{sec:hkks} and Eq.~(26) of Ref. \cite{herring}) is $T_s=
\frac {\hbar^2}{2m}\int \sum_{i=1}^N |\nablab\phi_i|^2 d^3\rr$. In
Eq.~(27) of Ref. \cite{herring}, Herring then defines angular
variables representing points on the surface of an $N$-dimensional
unit sphere as (in our notation) $u_i(\rr) =\phi_i/\rho^{1/2}$. In
Eq.~(28) of Ref. \cite{herring}, he then expresses the
noninteracting kinetic energy as $T_s = T_\theta + T_W$ where
$T_\theta$, which is dependent on the $u_i$, is termed the
relative-phase energy. Comparison of Eq.~(\ref{dec_Ts}) in this
paper and Eq.~(28) of Ref. \cite{herring} shows that (in the orbital
approximation) $T_{Cl}$ and $T_\theta$ are equivalent. Herring
interprets the relative-phase energy as the additional kinetic
energy resulting from the exclusion principle which requires the
$N$-electron phase to vary with position (when there is more than
one electron with the same spin). His results for a variety of
one-dimensional potentials show that $T_\theta$ is usually a
significant fraction of the kinetic energy and that $T_\theta$
generally becomes larger relative to $T_W$ as $Z$ increases.
\cite{herring} The contribution of the electron phase to the kinetic
energy, which is implicit in hydrodynamic formulations of quantum
mechanics, \cite{GD} has been noted in other contexts.
\cite{herring,Luo,DS2} For hydrogenic orbitals there is an explicit
relationship between the electron phase and the angular momentum and
for hydrogenic orbitals with nonzero angular momentum, $T_{Cl}$ is a
significant fraction of the kinetic energy (as shown below). If
hydrogenic orbitals are used as basis functions for the ground
states of multi-electron atoms then, as $Z$ increases, the exclusion
principle will force electrons into orbitals with higher angular
momentum and the number of electrons with a given angular momentum
will increase in a stepwise fashion. We note that this behavior has
been demonstrated for the Thomas-Fermi electron density
\cite{JL,Oliphant} and there have been several approaches which
include angular momentum effects in Thomas-Fermi theory.
\cite{Hellman,KN} In the work of Englert and Schwinger
\cite{ES1,ES2}, angular momentum effects are included for the
express purpose of correcting the Thomas-Fermi electron density for
large $r$.

Our expression for the noninteracting kinetic energy, $T_s = T_{Cl}
+ T_W$, is exact and requires no additional proof. However, to gain
insight into the nature of $T_{Cl}$, we now examine our expression
for basis functions that are the product of radial functions and
spherical harmonics (here the noninteracting kinetic energy is
simply the kinetic energy and Eq.~(\ref{dec_Ts}) becomes $T = T_{Cl}
+ T_W$). These basis functions are typically used to represent
one-electron orbitals for the ground states of multi-electron atoms.
For practical reasons they are usually Slater orbitals but, for
simplicity, we present results for hydrogenic orbitals. We
explicitly show that, for these basis functions, our expression for
the kinetic energy is correct and furthermore, that it is correct
for the radial distributions of the integrands of $T$, $T_{Cl}$ and
$T_W$. That is, that for each value of $r$, the integrand of $T$ is
equal to the sum of the integrands of $T_{Cl}$ and $T_W$. The
hydrogenic orbitals, $\psi{(n,l,m)}$, are dependent on the principal
quantum number $n$, the angular momentum quantum number $l$ and the
magnetic quantum number $m$ but the total energy is dependent only
on $n$ and is (in atomic units) $E$ = -1/2$n^2$. Then, from the
virial expression for Coulombic systems, the kinetic energy is $T$ =
-$E$ = 1/2$n^2$. The classical kinetic energy is zero for
$\psi{(2,0,0)}$ and $\psi{(2,1,0)}$ and, from direct calculation,
$T_W$ is 1/8 which is equal to $T$. However, the classical kinetic
energy is nonzero for $\psi{(2,1,1)}$ and $\psi{(2,1,-1)}$ and, from
direct calculation, both $T_{Cl}$ and $T_W$ are 1/16 and $T_{Cl}+
T_W$ is equal to $T$. Radial distributions (integrated over the
angular variables) of the integrands for $T_{Cl}$, $T_W$ and $T$ are
shown in Fig. 1(a). The radial distribution for $T_{Cl}$ is
dependent on $n$, $l$ and $|m|$ but the classical kinetic energy is
dependent only on $n$ and $|m|$ and $T_{Cl}$ = $\frac {|m|}{n}T =
|m|/2n^3$. Thus $T_{Cl}$ is constant for $n$ and $|m|$ fixed and
this is illustrated in Fig. 1(b)-(d) which shows the radial
distributions for $T_{Cl}$, $T_W$ and $T$ for $n$ = 5, $|m|$ = 1 and
$l$ = 1 to 3. In these three cases the radial distributions for
$T_{Cl}$ all integrate to 1/250. For $n$ and $l$ fixed, $T_{Cl}$
increases from 0 to $l/2n^3$ as $|m|$ increases from 0 to $l$ and
this is illustrated in Fig. 1(e)-(h) which shows the radial
distributions for $T_{Cl}$, $T_W$ and $T$ for $n$ = 5, $l$ = 4 and
$|m|$ = 1 to 4. In these four cases the radial distributions for $T$
are identical and in each of Fig. 1(b)-(h) the radial distributions
for $T$ integrate to 1/50.

For the ground states of multi-electron atoms we expect that
$T_{Cl}$ will be greater than zero but smaller than $T_{TF}$ (when
the reduced gradient is small $T_{TF}$ has been shown
\cite{herring,DS2} to be an upper bound to $T_{Cl}$) and, across a
row of the periodic table, $T_{Cl}$ generally increases as $Z$
increases. For example, the one-electron orbital for the ground
state of the C atom will have a larger $l$ = 1 contribution than
will that for the ground state of the Be atom. Correspondingly,
$T_s$ for the C atom will have a larger $T_{Cl}$ component than will
that for the Be atom. However, we have no algorithm for optimizing
the $T_{Cl}$ component of the one-electron orbital. Since $T_{Cl}$
is dependent on the $m$ value for each basis function this algorithm
would be, at best, order $N^3$ and could have no practical advantage
over the standard Kohn-Sham algorithm.

It is important to note that, in our approach, the classical
kinetic energy is zero if the orbital is real. Thus, whereas
$T_{Cl}$ is nonzero for $\psi{(2,1,1)}$ and $\psi{(2,1,-1)}$ (with
$|m|$=1), it is zero for the familiar $p_x$ and $p_y$ orbitals
(formed from their linear combinations). For these real orbitals
$T_W$ is 1/8 which is equal to $T$ and this is appropriate as,
although $m$ is not zero, the expectation value of $L_z$ is. To
obtain an expression corresponding to $T_s$ = $T_{Cl}+ T_W$ it is
necessary to partition the Weizs\"{a}cker term as $T_W$ =
$T_W^\phi$ + $T_W^{r,\theta}$ where $T_W^\phi$ results from local
fluctuations in $\phi$ (and corresponds to $T_{Cl}$ for $|m|$ = 1)
and $T_W^{r,\theta}$ results from local fluctuations in $r$ and
$\theta$ (as does $T_W$ for $|m|$ = 1, to which it is identical).
For the $p_x$ and $p_y$ orbitals the radial distributions of
$T_W^\phi$ and $T_W^{r,\theta}$ are identical to those of $T_{Cl}$
and $T_W$ in Fig. 1(a). From a practical viewpoint the expressions
$T_{Cl}+ T_W$ and $T_W^\phi$ + $T_W^{r,\theta}$ are completely
equivalent and are equally useful as decompositions of $T_s$ but
their interpretation is different.

\begin{figure*}[htbp]
\begin{center}
\includegraphics[width=\linewidth]{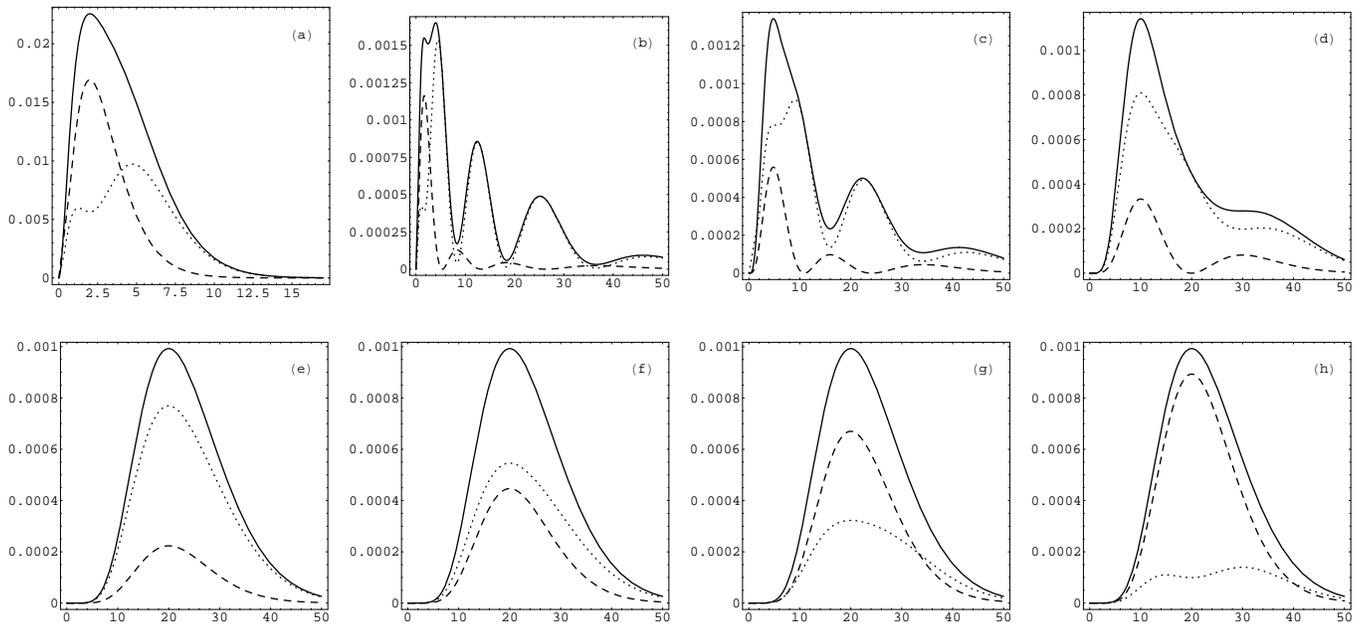}
\caption{Radial distributions (integrated over the angular
variables) of the integrands for $T_{Cl}$ (dashed curve), $T_W$
(dotted curve) and $T$ = $T_{Cl}$ + $T_W$ (solid curve) for
hydrogenic orbitals with (a) $n$ = 2, $l$ = 1, $|m|$ = 1; (b)-(d)
$n$ = 5, $|m|$ = 1 and $l$ = 1 to 3; (e)-(h) $n$ = 5, $l$ = 4 and
$|m|$ = 1 to 4. The horizontal axis is in atomic units.}
\label{fig:fig1}
\end{center}
\end{figure*}

\section{Conclusions}

In this paper we employed a recently formulated dequantization
procedure to obtain an exact expression for the kinetic energy which
is applicable to all kinetic-energy functionals. In this expression
the kinetic energy of an $N$-electron system is written as the sum
of the $N$-electron classical kinetic energy and the $N$-electron
purely quantum kinetic energy arising from the quantum fluctuations
that turn the classical momentum into the quantum momentum. Our
dequantization procedure also results in a term (identical to the
osmotic momentum of Nelson \cite{nelson}) which, when added to the
classical momentum results in the quantum momentum. We thereby
established a connection between Nelson's stochastic approach to
quantum mechanics, the Weizs\"{a}cker term and the Fisher Information
Theory. Moreover, the connection to Fisher Information Theory
provides a basis for an interesting conceptual interpretation of some
terms contributing to the kinetic-energy functional.

We wrote the $N$-electron purely quantum kinetic energy as the sum
of the (one-electron) Weizs\"{a}cker term which results from the
local quantum fluctuations and a kinetic correlation term which
results from the nonlocal quantum correlations. We also wrote the
$N$-electron classical kinetic energy as the sum of the
(one-electron) classical kinetic energy and another kinetic
correlation term. We then obtained an expression for the
noninteracting kinetic energy as the sum of the classical kinetic
energy and the Weizs\"{a}cker term. The Weizs\"{a}cker term is
well-known and the classical kinetic energy is related to the
Thomas-Fermi term which is also well-known. However, we believe that
our derivation, which obtains {\it both} these terms within a {\it
single} theoretical framework, is novel. Also, there are significant
differences between the classical kinetic energy and the
Thomas-Fermi term and we conclude with some further remarks on our
expression for the noninteracting kinetic energy.

Our expression is exact and we have shown that for the ground state
of the hydrogen atom it correctly reduces to the Weizs\"{a}cker term
while for the uniform electron gas it correctly reduces to the
Thomas-Fermi term (which is identical to the classical kinetic
energy for this system). However, the classical kinetic energy
(unlike the Thomas-Fermi term) is explicitly dependent on the
electron phase. The expression $T_s = T_{Cl} + T_W$ is therefore at
best order $N^3$ and can have no practical advantage over the
standard Kohn-Sham expression. To gain insight into the nature of
$T_{Cl}$, we examined our expression for basis functions that are
the product of radial functions and spherical harmonics and
established a direct connection between the classical kinetic energy
and the angular momentum. We believe that this intrinsic connection
between the angular momentum and a component of the noninteracting
kinetic energy is of significant conceptual value in showing the
information that should be incorporated in any kinetic-energy
functional.

For small and intermediate $Z$ atoms, the basic problem with the
expression $T_s = T_{TF} + \lambda T_W$ (or $\lambda T_{TF} + T_W$)
is that $T_W$ incorporates exactly a part of the noninteracting
kinetic energy that is also incorporated approximately in $T_{TF}$.
\cite{absp} This component of $T_{TF}$ should be removed and that is
why simply optimizing $\lambda$ offers only limited improvement.
\cite{CCH} The expression $T_s = T_{Cl} + T_W$ is a significant
improvement in this regard as $T_{Cl}$ and $T_W$ are completely
independent. However, as the classical kinetic energy is explicitly
dependent on the electron phase, our expression is manifestly not
orbital-free. As all explicit information regarding the electron
phase is lost in constructing the electron density it is clear that
any direct functional of $\ro$ which embodies this information must
be highly nonlocal. \cite{herring,herring2,LG,GAC,NHM}
Reconstructing this information from the electron density represents
a significant challenge for the development of accurate orbital-free
kinetic-energy functionals.

For large $Z$ atoms, the electron density is slowly varying for the
bulk of the electrons in the intermediate $r$ region and a
second-order gradient expansion is valid. However, this expression
is not valid for large $r$. Unfortunately, the large $r$ region is
(by virtue of the valence electrons) responsible for chemical
bonding and Thomas-Fermi theory cannot describe molecular systems.
Our expression is equally valid for intermediate and large $r$ but
it is much more difficult to evaluate. For large $Z$ atoms (where
the order $N^3$ aspect is of greatest concern) it would, in
principle, be possible to develop a hybrid approach in which
$T_{TF}$ + $\frac{1}{9}$$T_W$ (with the Dirac exchange functional
\cite{dirac}) is employed for the bulk of the electrons in the
intermediate $r$ region and corrected for large $r$ by evaluating
$T_{Cl}+T_W$ for the valence electrons only.

\acknowledgments

IPH acknowledges funding from NSERC and thanks Wilfrid Laurier University
for support. RAM acknowledges FAPESP for financial support.


\begin{thebibliography}{99}

\bibitem{YangParr}
R. G. Parr and W. Yang, \textit{Density Functional Theory of Atoms
and Molecules}, (Oxford University Press, New York, 1989).

\bibitem{DG}
R. M. Dreizler and E. K. U. Gross, \textit{Density Functional
Theory: An Approach to the Quantum Many Body Problem},
(Springer-Verlag, Berlin, 1990).

\bibitem{KH}
W. Koch and M. C. Holthausen, \textit{A Chemist's Guide to Density
Functional Theory}, (Wiley-VCH, Weinheim, 2000).

\bibitem{weizsacker}
C. F. v Weizs\"{a}cker, Z. Phys. {\bf 96}, 431 (1935).

\bibitem{nelson} E. Nelson, Phys. Rev. {\bf 150}, 1079 (1966);
E. Nelson, \textit{Dynamical Theories of Brownian Motion}
(Princeton Univ. Press, Princeton, 1967).

\bibitem{FE} I. F\'enyes, Z. Physik {\bf 132}, 81 (1952).

\bibitem{WE} W. Weizel, Z. Physik {\bf 134}, 264 (1953); {\bf 135}, 270 {1953}; {\bf 136}, 582 (1954).

\bibitem{MHD06}
R. A. Mosna, I. P. Hamilton and L. Delle Site, J. Phys. A {\bf
39}, L229 (2006).

\bibitem{thomas}
L. H. Thomas, Proc. Camb. Phil. Soc. {\bf 23}, 542 (1927).

\bibitem{fermi}
E. Fermi, Rend. Accad. Lincei {\bf 6}, 602 (1927).

\bibitem{MY}
N. H. March and W. H. Young, Proc. Phys. Soc. {\bf 72}, 182
(1958).

\bibitem{absp}
P. K. Acharya, L. J. Bartolotti, S. B. Sears, and R. G. Parr,
Proc. Nat. Acad. Sci. {\bf 77}, 6978 (1980).

\bibitem{GL} J. L. G\'azquez and E. V. Lude\~na, Chem. Phys. Lett.
{\bf 83}, 145 (1981).

\bibitem{gr}
J. L. G\'azquez and J. Robles, J. Chem. Phys. {\bf 76}, 1467
(1982).

\bibitem{acharya} P. K. Acharya, J. Chem. Phys. {\bf 78}, 2101
(1983).

\bibitem{KP} A. S. Kompaneets and E. S. Pavlovski, Sov. Phys.-JETP {\bf 4}, 328 (1957).

\bibitem{kirzhnits}
P. A. Kirzhnits, Sov. Phys.-JETP {\bf 5}, 64 (1957).

\bibitem{yang}
W. Yang, Phys. Rev. A {\bf 34}, 4575 (1986).

\bibitem{herring} C. Herring, Phys. Rev. A {\bf 34}, 2614 (1986).

\bibitem{Scott}
J. M. C. Scott, Philos. Mag. {\bf 43}, 859 (1952).

\bibitem{dirac}
P. A. M. Dirac, Proc. Cambridge Philos. Soc. {\bf 26}, 376 (1930).

\bibitem{Spruch}
L. Spruch, Rev. Mod. Phys. {\bf 63}, 151 (1991).

\bibitem{fisher}
R. A. Fisher, Proc. Cambridge Philos. Soc. {\bf 22}, 700 (1925).

\bibitem{nagy}
A. Nagy, J. Chem. Phys. {\bf 119}, 9401 (2003).

\bibitem{HK}
P. Hohenberg and W. Kohn, Phys. Rev. {\bf 136}, B864 (1964).

\bibitem{KS}
W. Kohn and L. J. Sham, Phys. Rev. {\bf 140}, A1133 (1965).

\bibitem{CCH}
G. K. Chan, A. J. Cohen and N. C. Handy, J. Chem. Phys. {\bf 114},
631 (2001).

\bibitem{IEMS}
S. S. Iyengar, M. Ernzerhof, S. N. Maximoff and G. E. Scuseria,
Phys. Rev. A {\bf 63}, 052508 (2001).

\bibitem{TW}
F. Tran and T. A. Weso{\l}owski, Chem. Phys. Lett. {\bf 360}, 209
(2002).

\bibitem{SLBB}
E. Sim, J. Larkin, K. Burke, C. W. Bock, J. Chem. Phys. {\bf 118},
8140 (2003).

\bibitem{JY}
H. Jiang and W. T. Yang, J. Chem. Phys. {\bf 121}, 2030 (2004).

\bibitem{CW}
J. D. Chai and J. A. Weeks, J. Phys. Chem. B {\bf 108}, 6870 (2004).

\bibitem{BC}
X. Blanc X and E. Cances, J. Chem. Phys. {\bf 122}, 214106 (2005).

\bibitem{ON}
I. V. Ovchinnikov and D. Neuhauser, J. Chem. Phys. {\bf 124}, 024105
(2006).

\bibitem{ZW} B. Zhou and Y. A. Wang, J. Chem. Phys. {\bf 124}, 081107 (2006).

\bibitem{MHD05}
R. A. Mosna, I. P. Hamilton and L. Delle Site, J. Phys. A {\bf 38},
3869 (2005), quant-ph/0504124.

\bibitem{Goldstein}
H. Goldstein, \textit{Classical Mechanics, 2nd ed.} (Addison-Wesley, Reading, MA, 1980).

\bibitem{Holland}
P. R. Holland, \textit{The Quantum Theory of Motion} (Cambridge
University Press, Cambridge, 1993).

\bibitem{sears}
S. B. Sears, R. G. Parr and U. Dinur, Isr. J. Chem. {\bf 19}, 165
(1980).

\bibitem{GD} S. K. Ghosh and B. M. Deb, Phys. Rep. {\bf 92}, 1
(1982).

\bibitem{Luo} S. Luo, J. Phys. A {\bf 35}, 5181 (2002).

\bibitem{DS2} L. Delle Site, J. Phys. A {\bf 38}, 7893 (2005).

\bibitem{JL} J. H. D. Jensen and J. M. Luttinger, Phys. Rev. {\bf 86}, 907 (1952).

\bibitem{Oliphant} T. A. Oliphant, Jr., Phys. Rev. {\bf 104}, 954 (1956).

\bibitem{Hellman}
H. Hellman, Acta Physicochem USSR {\bf 4}, 225 (1936).

\bibitem{KN}
G. Kemister and S. Nordholm, J. Chem. Phys. {\bf 76}, 5043 (1982).

\bibitem{ES1}
B.-G. Englert and J. Schwinger, Phys. Rev. A {\bf 29}, 2339 (1984).

\bibitem{ES2}
B.-G. Englert and J. Schwinger, Phys. Rev. A {\bf 32}, 47 (1985).


\bibitem{miao} M. S. Miao, J. Phys. A {\bf 34}, 8171 (2001).

\bibitem{ayers} P. W. Ayers, J. Math. Phys. {\bf 46}, 062107
(2005).

\bibitem{DS1} L. Delle Site, J. Phys. A {\bf 39}, 3047 (2006).

\bibitem{gadre} S. R. Gadre, Adv. Quantum Chem. {\bf 22}, 1 (1991).

\bibitem{romera1} E. Romera and J. S. Dehesa, Phys. Rev. A {\bf 50}, 256 (1994).

\bibitem{shannon} C. E. Shannon, Bell Syst. Tech. J. {\bf 27}, 623 (1948).

\bibitem{romera2}
E. Romera and J. S. Dehesa, J. Chem. Phys. {\bf 120}, 8906 (2004).

\bibitem{sagar} R. P. Sagar and N. L. Guevara, J. Chem. Phys. {\bf 123}, 044108 (2005).

\bibitem{herring2} C. Herring and M. Chopra, Phys. Rev. A {\bf 37}, 31 (1988).

\bibitem{LG} D. J. Lacks and R. G. Gordon, J. Chem. Phys. {\bf 100},
4446 (1994).

\bibitem{GAC} P. Garc\'ia-Gonz\'alez, J. E. Alvarellos and E. Chac\'on, Phys. Rev. A {\bf 54}, 1897 (1996).

\bibitem{NHM} N. H. March, Int. J. Quantum Chem. {\bf 92}, 1(2003).

\end{thebibliography}
\end{document}